# Optimizing Point-to-Multipoint Transmissions in High Speed Packet Access Networks

G. Araniti, V. Scordamaglia, A. Molinaro, A. Iera, G. Interdonato, and F. Spanò

*Abstract*— In this paper an innovative Radio Resource Management (RRM) algorithm is proposed with the purpose of increasing High Speed Packet Access (HSPA) performances, in terms of system capacity and service quality, when the Multimedia Broadcast Multicast Services (MBMS) is supplied. The proposed RRM algorithm exploits channel quality indications to set up point-to-multipoint connections to subgroups of multicast users and to select the proper modulation and coding schemes on the downlink. The number of subgroups is determined through an optimization technique that also takes into account the user satisfaction. An exhaustive simulation campaign is conducted to compare the proposed algorithm with the most promising approaches in the literature. Comparisons aim to assess the capability of the proposed RRM algorithm to efficiently manage group oriented services by providing an increment in terms of user satisfaction.

## I. INTRODUCTION

In the last few years the mobile telecommunications market has been characterized by a huge development that has exceeded even the most optimistic expectations. In such a context, *mobile broadcasting* is surely one of the most promising areas, which is expected to make a set of *value-added services* accessible from the mobile devices of groups of users (*group-oriented services*). Significant examples of these services are *mobile TV* [1], news services (e.g., e-press, sport, weather forecasts), games and, in general, downloading of multimedia contents (e.g., music, software, video).

The growing demand for *group oriented services* has led to a rapid evolution in the third generation (3G) mobile systems' architecture. The Universal Mobile Telecommunications Systems (*UMTS*) [2] can be considered as the first real step toward the convergence of broadband and mobile systems. Thanks to UMTS, mobile operators have started to provide multimedia services over cellular phones, such as video clips from sport events or live TV programs, video conferencing and file downloading. However, cost and technical limitations make UMTS unfit for satisfying the increasing demand for high speed data access. Just to give an example, with the limited *UMTS* network capacity, only low-resolution short video clips (e.g., 2 minutes at 128 kbps) [3] can be supported.

High-Speed Packet Access (*HSPA*) [4], standardized as part of Release 5, improves the UMTS network capacity by granting higher data rates in both uplink (High-Speed Uplink Packet Access, *HSUPA*) and downlink (High-Speed Downlink Packet Access, *HSDPA*) directions. HSDPA introduces a new transport channel, the High Speed-Downlink Shared Channel (HS-DSCH), which enables a theoretical maximum data rate of 14.4 Mbps. It can be shared among several users and handled through adaptive mechanisms based on the channel quality. Even if HSPA allows more efficient and flexible radio resource utilization, yet it is optimized for *unicast* services delivered over dedicated point-to-point (*PtP*) connections.

The diffusion of *group oriented services* has originated the need for delivering the same multimedia content from one sender to several receivers by exploiting point-to-multipoint (*PtM*) connections. In order to transmit efficiently the same content to several users at the same time, the 3G standard (Rel. 6) has been enhanced with Multimedia Broadcast Multicast Services (MBMS) [5], which provides integration of multicast and broadcast technologies into the existing 3G networks. MBMS supports *group-oriented services* in the UMTS Radio Access Network (UTRAN) by employing both *PtP* and *PtM* transport channels.

*PtP* transmissions can maximize the quality of service (QoS) offered to each user at the expense of an inefficient use of radio resources and the consequent reduction of the overall system capacity, especially in the presence of a large number of multicast users. In contrast, a *PtM* broadcast transmission "theoretically" provides unlimited capacity, with the disadvantage, however, to provide QoS on a group-basis instead of a single user basis. So, QoS will be bounded by the worst channel conditions with a resulting increase in "user dissatisfaction".

Efforts of the scientific community have been directed to find a trade-off between the system capacity and the QoS (defined in terms of data rate, block error rate, etc.).

The research work presented in this paper intends to give a contribution in this context. It indeed, proposes an innovative Radio Resource Management (RRM) algorithm with the purpose of increasing HSPA performances, in terms of system capacity and service quality, when the MBMS service is supplied. The proposed RRM policy is based on the idea of

The authors are with the DIMET Dept. - Università Mediterranea di Reggio Calabria, Loc. Feo di Vito – Reggio Calabria 89100 – Italy (contact author's e-mail:: araniti@unirc.it).

splitting a MBMS multicast group into subgroups, based on channel quality feedbacks from the users, and applying an optimization procedure to identify the subgroup division that gives the best performance in terms of user satisfaction. The proposed policy supports a subgroup-based adaptive modulation and coding mechanism rather than one based on single user.

The remainder of this paper is organized as follows. In Section II a brief state of the art of the research on MBMS in HSDPA networks is provided. In Section III the proposed RRM algorithm is described; results of a wide simulation campaign that compares the proposed policy with two reference algorithms are reported in Section IV. Concluding remarks and future works are summarized in Section V.

## II. MULTICAST TRANSMISSIONS IN HSPA NETWORKS

The MBMS standard provides the following options to support *group-oriented services*: the *PtP* Dedicated Channel (DCH), the *PtM* Forward Access Channel (FACH), the *PtP* or *PtM* High Speed-Downlink Shared Channel (HS-DSCH).

Several RRM policies have been proposed in the past to select the transport channel type that maximizes the number of MBMS users in the system [6-10]. Studies demonstrated that the *PtM* transmission over HS-DSCH is the most promising approach [10]. HS-DSCH enables efficient and flexible radio resource allocation, channel sharing from multiple users, and Adaptive Modulation and Coding (AMC) support [4]. The modulation and coding scheme (MCS) can be adjusted every Transmission Time Interval (TTI)[1] on the basis of periodic feedbacks on the channel quality provided by the User Equipments (UEs). The Channel Quality Indicator (*CQI*) is exploited by Node B (the UMTS base station) to select the MCS to use for data transmission to the UE in the next *TTI* [11]. Among the values of the CQI feedback cycle allowed by the standard specifications [12], in this paper CQIs are assumed to be forwarded to Node B every *20* TTIs.

*CQI* values depend on the maximum tolerated *Block Error Rate* (BLER), the distance between UE and Node B, the channel conditions in terms of *Signal to Interference Noise Ratio* (SINR), the UE category. In Table I the CQI/data-rate mapping is shown for Category 10 UEs.

The CQI feedback procedure well supports *PtP* traffic over HS-DSCH; but it cannot be straightforward applied to *group-oriented services*. In this case, in fact, potential receivers could have different CQIs, and then require different MCSs.

In [9] the authors proposed to provide *PtM* transmissions on HS-DSCH. They demonstrated that one High Speed-Physical Downlink Shared Channel (HS-PDSCH) can be shared among multicast users located in the same cell and interested in the same service. One physical channel can be shared only if one MCS is used for *PtM* transmission to all the receivers [9-11], [12]. The MCS is selected to ensure correct reception to the UE with the *worst* channel conditions (generally the minimum *CQI* is chosen to ensure 95% of the coverage). The proposed policy (hereinafter *Single Group*) allowed to drastically reduce the blocking probability of multicast calls compared to *PtP* transmissions. Nevertheless, based on this policy, UEs with higher *CQI* values perceive significant QoS degradation.

TABLE I
CQI MAPPING [11]

| CQI Values | Modulation | Transport Block Size | Codes Number | Single Channel Bit Rate (kbps) | Total Bit Rate (kbps) | $b_{code}$ | Code Rate | Data Rate (kbps) |
|---|---|---|---|---|---|---|---|---|
| 1 | QPSK | 137 | 1 | 480 | 480 | 960 | 0,1427 | 68,50 |
| 2 | QPSK | 173 | 1 | 480 | 480 | 960 | 0,1802 | 86,50 |
| 3 | QPSK | 233 | 1 | 480 | 480 | 960 | 0,2427 | 116,50 |
| 4 | QPSK | 317 | 1 | 480 | 480 | 960 | 0,3302 | 158,50 |
| 5 | QPSK | 377 | 1 | 480 | 480 | 960 | 0,3927 | 188,50 |
| 6 | QPSK | 461 | 1 | 480 | 480 | 960 | 0,4802 | 230,50 |
| 7 | QPSK | 650 | 2 | 480 | 960 | 960 | 0,3385 | 325,00 |
| 8 | QPSK | 792 | 2 | 480 | 960 | 960 | 0,4125 | 396,00 |
| 9 | QPSK | 931 | 2 | 480 | 960 | 960 | 0,4849 | 465,50 |
| 10 | QPSK | 1262 | 3 | 480 | 1440 | 960 | 0,4382 | 631,00 |
| 11 | QPSK | 1483 | 3 | 480 | 1440 | 960 | 0,5149 | 741,50 |
| 12 | QPSK | 1742 | 3 | 480 | 1440 | 960 | 0,6049 | 871,00 |
| 13 | QPSK | 2279 | 4 | 480 | 1920 | 960 | 0,5935 | 1139,50 |
| 14 | QPSK | 2583 | 4 | 480 | 1920 | 960 | 0,6727 | 1291,50 |
| 15 | QPSK | 3319 | 5 | 480 | 2400 | 960 | 0,6915 | 1659,50 |
| 16 | 16QAM | 3565 | 5 | 960 | 4800 | 1920 | 0,3714 | 1782,50 |
| 17 | 16QAM | 4189 | 5 | 960 | 4800 | 1920 | 0,4364 | 2094,50 |
| 18 | 16QAM | 4664 | 5 | 960 | 4800 | 1920 | 0,4858 | 2332,00 |
| 19 | 16QAM | 5287 | 5 | 960 | 4800 | 1920 | 0,5507 | 2643,50 |
| 20 | 16QAM | 5887 | 5 | 960 | 4800 | 1920 | 0,6132 | 2943,50 |
| 21 | 16QAM | 6554 | 5 | 960 | 4800 | 1920 | 0,6827 | 3277,00 |
| 22 | 16QAM | 7168 | 5 | 960 | 4800 | 1920 | 0,7467 | 3584,00 |
| 23 | 16QAM | 9719 | 7 | 960 | 6720 | 1920 | 0,7231 | 4859,50 |
| 24 | 16QAM | 11418 | 8 | 960 | 7680 | 1920 | 0,7434 | 5709,00 |
| 25 | 16QAM | 14411 | 10 | 960 | 9600 | 1920 | 0,7506 | 7205,50 |
| 26 | 16QAM | 17237 | 12 | 960 | 11520 | 1920 | 0,7481 | 8618,50 |
| 27 | 16QAM | 21754 | 15 | 960 | 14400 | 1920 | 0,7553 | 10877,00 |
| 28 | 16QAM | 23370 | 15 | 960 | 14400 | 1920 | 0,8115 | 11685,00 |
| 29 | 16QAM | 24222 | 15 | 960 | 14400 | 1920 | 0,8410 | 12111,00 |
| 30 | 16QAM | 25558 | 15 | 960 | 14400 | 1920 | 0,8874 | 12779,00 |

An interesting approach to overcome this problem has been proposed in [10], hereinafter named *Group-Based PtM (GB PtM)*. A multicast group that has joined a given Node B is split into *subgroups* based on their CQI values. Users in each *multicast subgroup* receive data with the same MCS; while different subgroups are assigned different MCSs. This is possible because the *AMC* mechanism is carried out on a subgroup basis rather than on individual user basis. Periodically, Node B collects all the CQI values via feedbacks from the users and defines the *subgroups* on the basis of the minimum CQI values of the subgroup members, based on equation (1) [10]:

$$Group\{CQI\} = Min[CQI_i], i \in N \qquad (1)$$

The number of subgroups is selected by Node B based on the most populated CQIs. Obviously, to serve all multicast users requiring a given service, the multicast subgroup with the lowest channel conditions is always activated, even if only one user is in such a situation.

The definition of the number of *multicast subgroups* that "maximizes system performance" is still an open issue. The purpose of this paper is to investigate such an issue. In particular, a RRM algorithm that adopts an optimization policy for subgroup formation is proposed with the aim to increase the capacity of the MBMS system and to maximize the user satisfaction.

---

[1] In HSDPA networks TTI has a duration of 2ms [4].

### III. THE PROPOSED RRM POLICY

The proposed RRM policy, hereinafter named *Enhanced-Group Based PtM (E-GB PtM)* foresees three phases:
1) *data collection*, during which Node B collects CQI feedbacks from each UE belonging to the multicast group of *N* users interested in the same content download;
2) *optimization and multicast subgroups creation*, where Node B groups the users into multicast subgroups according to the forwarded CQIs, determines the number of subgroups and selects the MCS for each subgroup (and not for a single user) in order to optimize the overall QoS;
3) *radio resource allocation* when the multicast service is finally provided.

Every 40 ms [11], the RRM procedure starts again from *phase 1* to check if a new optimization procedure is needed.

In this paper, an optimization algorithm is designed to manage *phase 2* with the aim of maximizing the number of multicast users served by Node B and, at the same time, the overall QoS.

The *optimization* procedure is based on the definition of a new QoS parameter, the *User Dissatisfaction Index (UDI)* $\omega_i$ denoted as follows:

$$\omega_i = \begin{cases} b_i - c_i & b_i \geq c_i \\ \infty & b_i < c_i \ \ OR \ \ c_i = 0 \end{cases} \quad (2)$$

where $b_i$ is the maximum data rate a user can support according to channel quality (in Table I), $c_i$ is the data rate assigned to a given *PtM* transmission. For a generic UE, the optimal condition is obtained when $b_i=c_i$. In this case $\omega_i = 0$ denotes full satisfaction of the *i-th* user. *UDI* approaches to infinity when $c_i=0$, or $b_i<c_i$. In the former case UE does not obtain any radio resources being in outage of multicast services, while, in the latter case, UE experiences an increase in terms of BLER with a consequence raise in terms of packets loss, because it receives multicast transmissions at a higher data rate than maximum allowable.

Under the assumption that all UEs with the same CQI belong to the same multicast subgroup, an overall QoS index called *Global Dissatisfaction Index (GDI)* $\Omega_i$ can be defined:

$$\Omega = \frac{1}{N} \sum_{k=1}^{N_{CQI}} \omega_k \cdot U_k \quad (3)$$

$U_k$ ( $\omega_k$ ) being the number (*UDI*) of Users associated to the *k-th* subgroup and $N_{CQI}$ the maximum number of considered CQI. This new overall QoS index takes into account the mean dissatisfaction value among all *N* UEs in the cell.

The problem of determining the optimal value of *GDI* can be recast into a constrained minimization problem. In order to formalize the minimization problem, the multicast subgroups configuration set $A = \{a_1, a_2 ... a_{N_{CQI}}\}$ is introduced. Generic element $a_k$ is a logical value: $a_k=1(0)$ means that data transmission with data rate $b_k$ is (is not) enabled.

If $B = \{b_1, b_2 ... b_{N_{CQI}}\}$ is the set of maximum data rate at varying of CQI, for a given set A, all UEs with the same *k-th* CQI level will receive a multicast service with an *assigned data rate* $c_k^A$ evaluated according to equation:

$$T_k(A) = \{b_j \in B : a_j = 1, j \leq k\} \quad (4)$$

$$c_k^A = \max T_k(A) \quad (5)$$

From (5), $c_k^A$ can be less or equal to $b_k$ and, specifically, $c_k^A = b_k$ only if $a_k = 1$. Moreover, given the *i-th* UE with CQI level *k*, the assigned data rate $c_i$ introduced in (2) will be equal to $c_k^A$. In view of this, it is worth noting that (3) strongly depends on set *A*.

Moreover it is introduced the following set

$$NoC = \{NoC_1, NoC_2 ... NoC_{N_{CQI}}\} \quad (6)$$

where the generic element $NoC_k$ is the number of channelization codes associated to *k-th* CQI level. Finally, from (6)

$$NoC(A) = \{NoC_k \in NoC : a_k = 1\} \quad (7)$$

$\sum NoC(A)$ represents the total amount of channelization codes utilized for all *PtM* transmissions.

The proposed AMC mechanism, necessary to optimize the fruition of the multicast service for all the subscriber user, can be recast as follows

$$\begin{aligned} \Pi &= \arg \min_A \Omega \\ s.t. & \\ \sum & NoC(A) \leq M \end{aligned} \quad (8)$$

$\Pi$ being the optimal configuration of set *A* and *M* the maximum allowable number of channelization codes for each TTI for all *PtM* transmissions.

**Remark 1**. If global amount of channelization codes is less than *M*, i.e.

$$\sum_{k=1}^{N_{CQI}} NoC_k \leq M \quad (9)$$

an optimal solution of (8) can be found by considering for each subgroup of UEs, an assigned data rate $c_k$ equal to the maximum data rate $b_k$. This implies that, for each UE, UDI and GDI are null.

**Remark 2**. Due to the fact that the number of possible configuration of $A$ is bounded ($2^{N_{CQI}}$), the minimization problem (8) admits one solution at least.

**Remark 3**. If (8) admits more than one solution, a reasonable choice to isolate one solution, can be any configuration of $A$ that minimizes the number of channelization codes allocated for *PtM* transmission, allowing to utilize the remaining resources to support other types of traffic.

## IV. SIMULATION AND RESULTS

The effectiveness of the proposed approach is assessed by simulations. Channel and system simulators have been implemented in Matlab/Simulink®. A HS-DSCH channel simulator allowed to evaluate the relation among *Signal to Interference Noise Ratio* (SINR), *Block Error Rate* (BLER), and CQIs. *SINR* has evaluated at the output of the receiver, using the following equation:

$$SINR = SF_{16} \frac{P_{HS-DSCH}}{P_{own}} \frac{1}{p + G^{-1}} \quad (10)$$

where $SF_{16}$ is the Spreading Factor equal to *16*, $P_{HS-DSCH}$ is the HS-DSCH transmission power, $P_{own}$ is the own cell interference, $p$ is the orthogonality factor (that can be zero in case of perfect orthogonality), and $G$ is the *geometry factor* defined as follows:

$$G = \frac{P_{own}}{P_{other} + P_{noise}} \quad (11)$$

where $P_{other}$ is the interference from neighboring cells and $P_{noise}$ is the *Additive White Gaussian Noise* (*AWGN*). The *geometry factor* indicates the distance from the base station; a lower *G-factor* is expected when a user is near the cell edge, where the interference from the neighboring cells is higher than the own cell interference.

The transmission channel has been modeled by taking into account path loss, shadowing, and multipath fading phenomena. Path loss has been modeled according to the *Okumura-Hata* model, shadowing through the *lognormal distribution*, and multipath fading in accordance to the 3GPP *ITU Pedestrian-A* model (3km/h) [13]. Furthers simulation assumptions are reported in Table II.

Fig. 1 shows SINR values when varying the CQI for different values of BLER (5%, 10%, 15%, and 20%). The SINR value estimated by a single UE influences the *CQI* selection that, in turn, affects the following parameters (please, refer to Table I): *(i)* modulation scheme (QPSK or 16 QAM); *(ii)* number of codes; *(iii)* code rate; *(iv)* data and bit rate; *(v)* transport block size, and so on. Obviously how the CQI is related to the listed parameters strictly depends on the nature of the UE. Indeed, values reported in Table I and utilized in the further simulation campaigns are related to UEs - Category 10 [11].

TABLE II – SIMULATION ASSUMPTIONS

| Parameter | Value |
|---|---|
| Cellular layout | Hexagonal grid |
| Cell radius | 550 m |
| Number of neighbouring cells | 18 |
| Maximum BS Tx power | 20 W |
| Base station antenna gain | 11,5 dBi |
| Other BS Tx power | 5 W |
| Common channel power | 1 W |
| HS-DSCH Used Power | 12 W |
| Orthogonality factor | 0,5 |
| Pathloss model | Okumura Hata |
| Shadow fading model | Lognormal distribution |
| Multipath fading | ITU Pedestrian A [13] |
| Thermal noise | -100 dBm |
| MBMS sessions | 1 |
| Avarage N. of Multicast UEs/cell | 100 |
| Maximum number of code | 15 |
| BLER target | 5-10-15-20% |
| Terminal Category | 10 |

In this preliminary study the effectiveness of the proposed *RRM* algorithm is assessed in a scenario characterized by different numbers of *UEs* (see Table II) uniformly spread within a reference area covered by *Node B*. It is the most general case in which UEs can experience a large range of *SINR*. As a consequence, the probability that all CQI values are experienced is very high.

Preliminary results obtained by proposed algorithm have been compared to those reported in [9] and [10] showing that the proposed optimization procedure allows reducing significantly the *GDI* and, consequently, improving the QoS perceived by each single user.

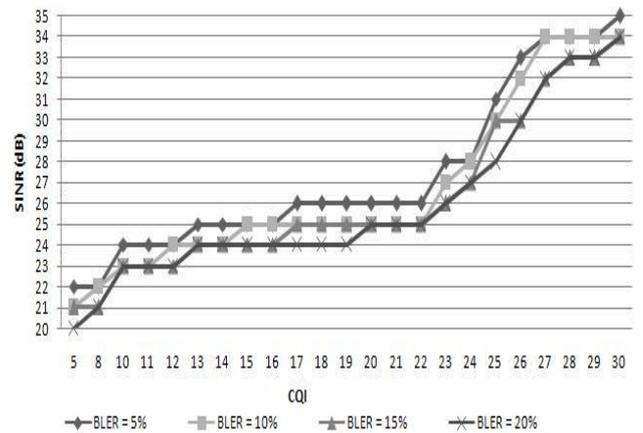

Fig. 1. SINR vs. CQIs for different values of BLER (5%, 10%, 15%, and 20%)

Fig. 2 depicts a snapshot showing the percentage number of users experiencing each value of CQI. This corresponds to the distribution of CQI values reported to a given Node B by UEs in a particular TTI. It is worth nothing that for a uniform scenario and under the assumptions reported in Table II, CQI assumes about all the possible values between *1* and *23*.

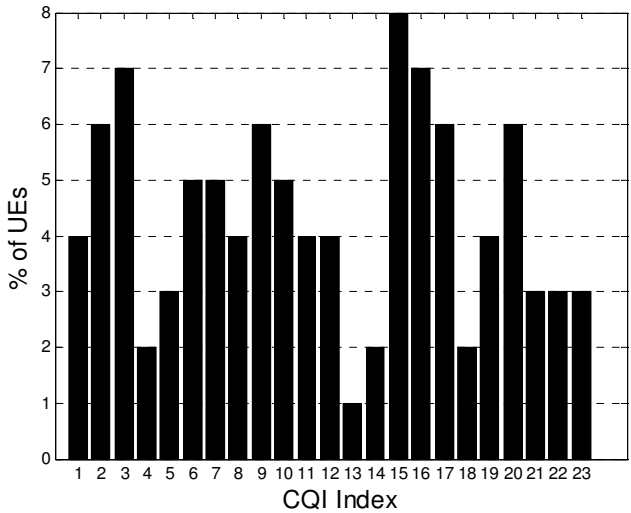

Fig. 2. Distribution of UEs with different CQI values

Fig. 3 shows how the three examined policies distribute users into different multicast subgroups. It is worth noting that the Single Group (*SG*) algorithm serves all multicast users with the lowest allowable *Data Rate*, while the other two policies generate multiple multicast subgroups with the purpose to increase the performance in term of *Data Rate*. The Group Based (*GB*) algorithm bases its choice on the most populous group of users [9] while the Enhanced Group Based (*E-GB*) algorithm takes its choices according to the optimization procedure proposed in this paper.

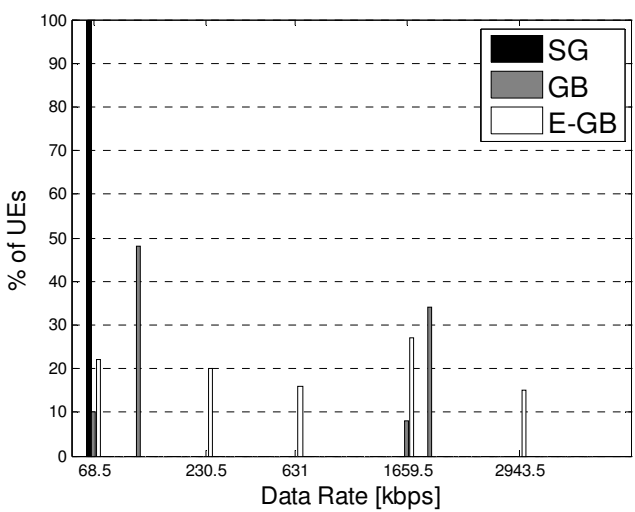

Fig. 3. Multicast Subgroups and relevant Data Rates

Fig. 4 shows GDI values normalized to the maximum allowable mean data rate. From this figure it clearly emerges that the proposed optimization algorithm introduces a significant QoS improvements for MBMS services.

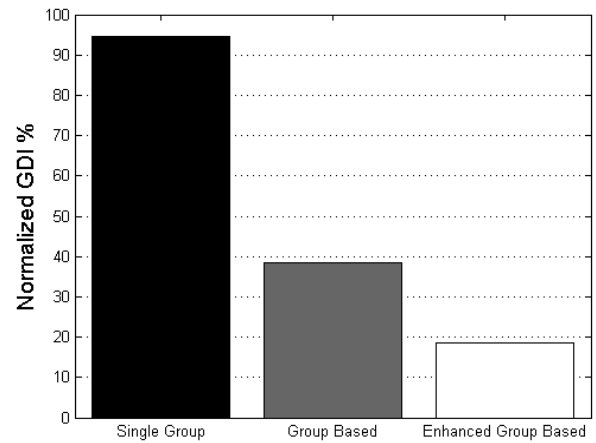

Fig. 4. Normalized Global Dissatisfaction Index

## V. CONCLUSION AND FUTURE WORKS

In this paper the multicast delivery features in 3G mobile networks with HSPA and MBMS have been augmented, in terms of system capacity and service quality, by using *PtM* transmissions over HS-DSCH with an appropriate RRM algorithm. An optimization procedure for maximizing user satisfaction has been proposed and validated. Obtained results showed that the proposed RRM policy offers significant QoS improvement to multicast users.